\documentclass[10pt,a4paper]{article}
\usepackage{amsmath}
\usepackage{amssymb,epsfig}

\begin{document}
\title{Harrison transformation of hyperelliptic solutions and 
charged dust disks
}
\author{C.~Klein, Max-Planck-Institut f\"ur Physik, F\"ohringer Ring 
6, 80805 M\"unchen, Germany        
}
\date{\today}    

\maketitle

\begin{abstract}
    We use a Harrison transformation on solutions to the 
    stationary axisymmetric Einstein equations 
    to generate solutions of the Einstein-Maxwell equations. The case 
    of  hyperelliptic solutions to the Ernst equation is studied in 
    detail. Analytic expressions for the metric and the multipole moments are 
    obtained. As an example we consider the transformation of a family 
    of counter-rotating dust disks. The resulting solutions can be 
    interpreted as disks with currents and 
    matter with a purely azimuthal pressure or as two streams of 
    freely moving charged particles. 
    We discuss interesting limiting cases as the extreme 
    limit where the charge becomes identical to the mass, and the 
    ultrarelativistic limit where the central redshift diverges.
\end{abstract}

\section{Introduction}
Electro-magnetic fields, especially magnetic fields are of 
astrophysical importance in the context of neutron stars, white dwarfs 
and galaxy formation. To get a complete relativistic understanding of 
such situations, one has to study the coupled Einstein-Maxwell equations. 
Since the stationary axisymmetric Einstein-Maxwell equations in vacuum 
in the form of Ernst \cite{ernst2}
are completely integrable (see \cite{maison79}), powerful solution 
generating techniques are at hand to obtain physically interesting 
solutions. But the equations in the matter region do not seem to be 
integrable, which makes it difficult to find 
global solutions which hold 
both in a three-dimensionally extended matter region  and in vacuum.

In the case of infinitesimally thin disks, the equations in the 
matter region reduce to ordinary differential equations the solutions 
of which 
determine boundary data for the vacuum equations. Such disks 
are discussed in astrophysics as models for Newtonian galaxies and the 
matter in accretion disks around black holes. Bi\v{c}\'ak, 
Lynden-Bell and Katz \cite{blk93} studied static spacetimes, from 
which they removed a strip and glued the remainder together at the 
equatorial plane. The non-continuous normal derivatives at the equatorial 
plane lead to a $\delta$-type energy-momentum tensor which can be 
interpreted as an infinitely extended disk made up of 
counter-rotating dust. This method was extended in \cite{ledvinka} to 
generate disk sources of the Kerr-metric. With the same techniques, 
disk sources for Kerr-Newman metrics \cite{zofka}, static axisymmetric 
spacetimes with magnetic fields \cite{letelier} and conformastationary 
metrics \cite{kbl99} were given. By construction all these disks have 
an infinite extension but finite mass.

In the present paper we give solutions to the Einstein-Maxwell 
equations with disk sources of finite extension. The solutions are 
obtained by exploiting the $SU(2,1)$-invariance of the stationary 
vacuum Einstein-Maxwell equations (see \cite{nk} to \cite{bgm}). We use a 
so-called Harrison transformation \cite{harrison} on disk 
solutions on hyperelliptic Riemann surfaces 
for the pure Einstein case, which are given in terms of Korotkin's 
\cite{korot88} theta functional solutions, 
to obtain charged disks with 
currents. In this case analytic formulas for the complete metric can 
be obtained. 
As an example we discuss the transformation of a family of 
counter-rotating disks \cite{prl2} which contains the Bardeen-Wagoner 
disk \cite{bawa} (see \cite{neugebauermeinel} for the exact solution)
as a limiting case. Following \cite{zofka} the resulting disk can 
be interpreted
either as a disk with a purely azimuthal pressure and 
currents or in 
a certain range of the physical parameters 
as two streams of freely moving charged particles, i.e.\ 
which move on electro-geodesics (solution to the 
geodesic equation in the presence of a 
Lorentz force). We discuss the energy-momentum tensor at the disk, 
the multipole moments and interesting limiting cases.

The paper is organized as follows. In section 2 we collect results on 
the stationary axisymmetric case in the absence of electromagnetic 
fields, solutions on hyperelliptic Riemann surfaces, and the family 
of disks discussed in \cite{prl2,prd4}. The Einstein-Maxwell equations, 
their group structure and the action of Harrison transformations on 
the multipoles are discussed in section 3. In section 4 we consider a 
Harrison transformation of the counter-rotating disks \cite{prl2} 
and discuss the metric and the energy-momentum tensor at the disk. 
The results are summarized in section 5 together with some concluding 
remarks.

\section{Stationary axisymmetric Einstein equations, theta functional 
solutions and counter-rotating dust disks}
In this section we will  summarize results on the integrability of the 
Ernst equation, a class of solutions on hyperelliptic 
Riemann surfaces \cite{korot88} and a member of this class 
describing counter-rotating dust disks which were discussed 
in \cite{prd3} and \cite{prd4}. 
\subsection{Ernst potential and metric}
We use the Weyl--Lewis--Papapetrou metric (see e.g.\
\cite{exac})
\begin{equation}
\mathrm{ d} s^2 =-e^{2U}(\mathrm{ d} t+a\mathrm{ d} \phi)^2+e^{-2U}
\left(e^{2k}(\mathrm{ d} \rho^2+\mathrm{ d} \zeta^2)+ \rho^2\mathrm{
d} \phi^2\right), \label{vac1}
\end{equation}
where $\rho$ and $\zeta$ are Weyl's canonical coordinates and
$\partial_{t}$ and $\partial_{\phi}$ are the two commuting
asymptotically timelike respectively spacelike Killing vectors.  With
$\xi=\zeta-i\rho$, we define the complex Ernst potential
$\mathcal{E}=f+ib$  
which is subject to the Ernst equation 
\cite{ernst1}
\begin{equation}
\mathcal{E}_{\xi\bar{\xi}}-\frac{1}{2(\bar{\xi}-\xi)}(
\mathcal{E}_{\bar{\xi}}-
\mathcal{E}_{\xi})=\frac{2
}{\mathcal{E}+\bar{\mathcal{E}}} \mathcal{E}_{\xi} 
\mathcal{E}_{\bar{\xi}} \label{vac10}\enspace,
\end{equation}
where a bar denotes complex conjugation in ${\mathbb C}$. If the 
Ernst potential is real, the corresponding spacetime is static and 
belongs to the so-called Weyl class. For a 
given Ernst potential the metric
(\ref{vac1}) follows from $e^{2U}=f$ and 
\begin{eqnarray}
     a_{\xi}&=&2\rho\frac{(\mathcal{E}-\bar{\mathcal{E}})_{\xi}}{
     (\mathcal{E}+\bar{\mathcal{E}})^{2}}\label{vac9},\\
k_{\xi}&=&(\xi-\bar{\xi}) \frac{\mathcal{E}_{\xi}\bar{\mathcal{E}}_{\xi}}{(
\mathcal{E}+\bar{\mathcal{E}})^{2}} \label{2.10a10}.
\end{eqnarray}
Thus the complete metric can be obtained 
from a given Ernst potential via quadratures.

Solutions with equatorial symmetry, i.e.\ a class of solutions 
where the metric functions have a reflection symmetry at the 
equatorial plane $\zeta=0$ (this implies $\mathcal{E}(-\zeta)=
\bar{\mathcal{E}}(\zeta)$), are of special physical interest. 
In a Newtonian setting it can be proven that perfect fluids 
in thermodynamical equilibrium 
lead to equatorially symmetric situations, and the same is assumed to 
hold in a general relativistic context. A consequence of this 
condition is that Newman-Unti-Tambourini-parameters are ruled out.

The stationary axisymmetric Einstein equations in vacuum were shown 
to be completely integrable in \cite{maison} and \cite{belzak}. 
This implies that the Ernst equation has an infinite 
dimensional symmetry group, the Geroch group (see \cite{breimai}).
The equation can be treated as the integrability condition for an 
overdetermined linear differential system for a $2\times 2$-matrix $\Psi$ 
(see \cite{korot92})
\begin{eqnarray}
    \Psi_{\xi}\Psi^{-1} & = & \left(
    \begin{array}{cc}
        \mathcal{M} & 0  \\
        0 & \mathcal{N}
    \end{array}
    \right)+\sqrt{\frac{K-\xi}{K-\bar{\xi}}}\left(
    \begin{array}{cc}
        0 & \mathcal{M}  \\
        \mathcal{N} & 0
    \end{array}\right)
    \nonumber \\
\Psi_{\bar{\xi}}\Psi^{-1} & = & \left(
    \begin{array}{cc}
        \bar{\mathcal{N}} & 0  \\
        0 & \bar{\mathcal{M}}
    \end{array}
    \right)+\sqrt{\frac{K-\xi}{K-\bar{\xi}}}\left(
    \begin{array}{cc}
        0 & \bar{\mathcal{N}}  \\
        \bar{\mathcal{M}} & 0
    \end{array}\right)
    \label{a1};
\end{eqnarray}
here $K$ is the so-called spectral parameter, and
\begin{equation}
    \mathcal{M}=\frac{\bar{\mathcal{E}}_{\xi}}{\mathcal{E}+\bar{\mathcal{E}}},
    \quad \mathcal{N}=\frac{\mathcal{E}_{\xi}}{\mathcal{E}+\bar{\mathcal{E}}}
    \label{MN}.
\end{equation}
The spectral 
parameter is defined on the family of Riemann surfaces $\Sigma_{0}$
given by the algebraic relation between $\mu_{0}$ and $K$,
\begin{equation}
    \mu_{0}^{2}=(K-\xi)(K-\bar{\xi})
    \label{mu0}.
\end{equation} 
Notice the special feature of the 
Ernst equation 
that the branch points $\xi$, $\bar{\xi}$ depend on the spacetime 
coordinates. 

It is a consequence of the linear system (\ref{a1}) that a 
matrix $\Psi$ considered as a function on $\Sigma_{0}$ and 
subject to the following conditions leads to a 
solution to the Ernst equation:\\
I. The matrix $\Psi$ is at least twice differentiable with respect to 
$\xi$ and $\bar{\xi}$. $\Psi(P)$ is holomorphic and invertible at the branch 
points $\xi$ and $\bar{\xi}$ such that the logarithmic derivative
$\Psi_{\xi}\Psi^{-1}$ has a pole at $\xi$ and 
$\Psi_{\bar{\xi}}\Psi^{-1}$ has a pole  at $\bar{\xi}$.\\
II. All singularities of $\Psi$ on $\Sigma_{0}$  
are such that the logarithmic derivatives 
$\Psi_{\xi}\Psi^{-1}$ and $\Psi_{\bar{\xi}}\Psi^{-1}$ are 
holomorphic there. 

III. $\Psi$ is subject to the reduction condition
\begin{equation}
	\Psi(P^{\sigma_{0}}) = \sigma_3 \Psi(P) \sigma_2
	\label{lin7},
\end{equation}
where  $\sigma_2$, $\sigma_{3}$ are Pauli matrices, and where 
$\sigma_{0}$ denotes the involution on $\Sigma_{0}$ which 
interchanges the sheets.\\
IV. The normalization and reality condition
\begin{equation}
	\Psi(P=\infty^+)=\left(
	\begin{array}{cc}
		\bar{\mathcal{E}} & -i  \\
		\mathcal{E} & i
	\end{array}
	\right)
	\label{lin9}.
\end{equation}
The function $\mathcal{E}$ in (\ref{lin9}) is a solution to the 
Ernst equation.

We note that the choice of $\sigma_{2}=\left(
\begin{array}{cc}
    0 & -i  \\
    i & 0
\end{array}\right)$ in condition III is a gauge condition, which 
does not fix the gauge completely, however. The remaining gauge 
freedom is due to matrices $C(K)=\kappa_{1}(K) \hat{1}+\kappa_{2}(K) 
\sigma_{2}$, where the $\kappa_{i}$ do not depend on $\xi$, 
$\bar{\xi}$ and obey the asymptotic conditions $\kappa_{1}(\infty)=1$ and 
$\kappa_{2}(\infty)=0$. The matrices $C$ act on $\Psi$ 
in the form $\Psi\to \Psi C(K)$. 
Since it is a consequence of (\ref{a1}) that $\det \Psi=F(K) e^{2U}$ 
where $F(K)$ 
is independent of $\xi$, $\bar{\xi}$ 
we can use this gauge freedom to 
choose $F(K)=1$.
The linear system (\ref{a1}) leads for the matrix $\chi(P)=
\Psi^{-1}(\infty^{-})\Psi(P)$ to one of the linear systems 
used in \cite{breimai}. This parametrization, especially
\begin{equation}
    \Psi^{-1}(\infty^{-})\Psi(\infty^{+})=\frac{1}{\mathcal{E}+
    \bar{\mathcal{E}}}\left(
    \begin{array}{cc}
        2\mathcal{E}\bar{\mathcal{E}} & 
        i(\bar{\mathcal{E}}-\mathcal{E})  \\
         i(\bar{\mathcal{E}}-\mathcal{E}) & 2
    \end{array}\right)
    \label{a3},
\end{equation}
reveals that the Ernst equation is invariant under  $SU(1,1)/SO(1,1)$ 
transformations. For more details on the group aspect see 
\cite{breimai} and the 
discussion in the next section.

For given  $\Psi(K)$, one can
directly determine the metric function $a$ without having to 
integrate relation (\ref{vac9}). We denote by $D_{P} F(P)$ the 
coefficient of the linear term in the expansion of the function 
$F$ in the local parameter near $P$. With (\ref{a1}) one finds for 
the matrix $\mathcal{S}=
\Psi^{-1}(\infty^{+})D_{\infty^{+}}\Psi(\infty^{+})$
\begin{equation}
   \mathcal{S}_{\xi}:= (\Psi^{-1}(\infty^{+})D_{\infty^{+}}\Psi)_{\xi}=\frac{\xi-
    \bar{\xi}}{2(\mathcal{E}+\bar{\mathcal{E}})^{2}}\left(
    \begin{array}{cc}
        (\mathcal{E}\bar{\mathcal{E}})_{\xi} & i(\bar{\mathcal{E}}-
	\mathcal{E})_{\xi}\\
        i(\mathcal{E}^{2}\bar{\mathcal{E}}_{\xi}-\bar{\mathcal{E}}^{2}
	\mathcal{E}_{\xi})& -(\mathcal{E}\bar{\mathcal{E}})_{\xi}
    \end{array}\right)
    \label{a4}.
\end{equation}
This implies with (\ref{vac9})
\begin{equation}
    a-a_{0}=-2\mathcal{S}_{12}=-2(\Psi^{-1}(\infty^{+})D_{\infty^{+}}\Psi)_{12},
    \label{a5}
\end{equation}
where $a_{0}$ is a constant which is fixed by the condition 
that $a$ vanishes on the regular part of the axis.

\subsection{Solutions on hyperelliptic surfaces}
Using the integrability of the Ernst equation, Korotkin 
\cite{korot92} gave solutions 
on hyperelliptic Riemann surfaces $\Sigma_{g}$ of genus $g$ which
are defined by the algebraic relation
\begin{equation}
    \mu^{2}=(K-\xi)(K-\bar{\xi})
\prod_{i=1}^{g}(K-E_{i})(K-\bar{E}_{i})
    \label{mu}.
\end{equation}
The points $E_{i}=\alpha_{i}-i\beta_{i}$
have to be constant with respect to 
the physical coordinates.  We order the branch points with $\mbox{Im}E_{i}<0$ in a way that 
$\Re E_{1}<\Re E_{2}<\ldots <\Re E_{g}$ and assume for 
simplicity that the real parts of the $E_{i}$ are all different.
On this surface we introduce a canonical basis of cycles $(a_{i}, 
b_{i})$, $i=1,\dots,g$,
the $g$ differentials of
the first kind $\mathrm{ d}\omega_i$ normalized 
by the condition $\oint_{a_i}\mathrm{ d}\omega_j=2\pi\mathrm{ i} 
\delta_{ij}$, and
the Abel map $\omega_i(P)=\int_{P_{0}}^{P}\mathrm{ d}\omega_i$.
Furthermore, we define
the Riemann matrix $\Pi$ with the elements $\pi_{ij}=
\oint_{b_i}\mathrm{ d}\omega_j$, and the theta function 
\begin{equation}
    \Theta\left[m\right](z)=
\sum_{N\in{Z}^g}^{}\exp\left\{\frac{1}{2}\left\langle\Pi (N+
\frac{m^{1}}{2}),(N+\frac{m^{1}}{2})
\right\rangle+\left\langle
(z+\pi i m^{2}),(N+\frac{\alpha}{2})
\right\rangle\right\}
    \label{theta}
\end{equation}
with half integer
characteristic $[m]=\left[m^{1}
\atop m^{2}\right]$ and $m^{1}_i,m^{2}_i=0,1$ 
($\left\langle N,z\right\rangle=\sum_{i=1}^g N_iz_i$). We write  
$\Theta$ for the theta 
function with characteristic $[0]$, and 
$P=(K,\pm \mu(K))$ or $K^{\pm}$  for a point 
$P\in \Sigma_{g}$ (the $+$ sheet will be defined by the condition 
that $\lim_{K\to\infty}\mu(K)/K^{g}=+1$ and similarly for the 
$-$-sheet). The 
normalized (all $a$--periods zero) differential of the third kind with 
poles at $P_{1}$ and $P_{2}$  and residue $+1$ and 
$-1$  respectively will be denoted by $\mathrm{ d} 
\omega_{P_{1}P_{2}}$.

Let $\ln G(K)$ be a H\"older-continuous function subject to the 
condition $G(\bar{K})=\bar{G}(K)$ and 
\begin{equation}
    I= \frac{1}{2\pi i}\int\limits_\Gamma \ln
G(\tau)d\omega_{\infty^{+}\infty^-}(\tau), \quad  
u_{i}=\frac{1}{2\pi
i}\int_{\Gamma}^{}\ln G d\omega_{i},
    \label{Iu}
\end{equation}
where  $\Gamma$ is the
covering of the imaginary axis in the +-sheet of $\Sigma_{g}$ between
$-i$ and $i$. Then one obtains solutions to the Ernst 
equation in the form
\begin{equation}
\mathcal{E}(\rho,\zeta)=\frac{\Theta[m](\omega(\infty^{+})+u
+\omega(\bar{\xi}))}{
\Theta[m](\omega(\infty^{+})-u+\omega(\bar{\xi}))}e^{I}
\label{rel1a}.
\end{equation}
Notice that these solutions depend only via the branch points of 
$\Sigma_{g}$ on the physical coordinates.  In \cite{prd2} it was 
shown that solutions of the above form on a Riemann surface of even 
genus $g=2s$ given by 
$\mu^{2}=(K-\xi)(K-\bar{\xi})\prod_{i=1}^{s}(K^{2}-E_{i}^{2})(K^{2}-
\bar{E}_{i}^{2})$ with a function $G$ subject to
$G(-K)=\bar{G}(K)$ lead to an equatorially symmetric Ernst potential.

The  Ernst potential (\ref{rel1a})
follows with (\ref{lin9}) and $J=\frac{1}{2\pi i}\int_{\Gamma}^{}\ln
G d
\omega_{PP^{\sigma}}$ from a matrix $\Psi$ of the form
\begin{equation}\label{a7}
    \Psi=e^{I/2}\sqrt{\frac{\det(\infty)}{\det(K)}}
    \left(
    \begin{array}{cc}
        \frac{\Theta[m](u+\omega(P))}{
           \Theta[m](u+\omega(\infty^{-}))}e^{J/2}
     & -i 
    \frac{\Theta[m](u+\omega(P^{\sigma}))}{
        \Theta[m](u+\omega(\infty^{-}))}e^{-J/2}
     \\
    \frac{\Theta[m](u+\omega(P)+
        \omega(\bar{\xi}))}{
        \Theta[m](u+\omega(\infty^{-})+\omega(\bar{\xi}))}e^{
	J/2}
        & i
   \frac{ \Theta[m](u+\omega(P^{\sigma})+\omega(\bar{\xi}))
       }{
       \Theta[m](u+\omega(\infty^{-})+\omega(\bar{\xi}))}e^{
       -J/2}
    \end{array}\right),
\end{equation}
with $\det(K) =\Theta[m](u+\omega(P))\Theta[m](u-\omega(P)+
    \omega(\bar{\xi}))+\Theta[m](u-\omega(P))\Theta[m](u+\omega(P)+
    \omega(\bar{\xi}))$.

To determine the metric function $a$ via (\ref{a5}), we have to 
calculate the matrix 
$\mathcal{S}$ which leads with 
(\ref{a7}) to 
\begin{equation}
    \mathcal{S}_{12}=\frac{1}{2if}D_{\infty^{+}}\ln\frac{\Theta(u+\omega(\infty^{-}))
    }{\Theta(u+\omega(\infty^{-})+\omega(\bar{\xi}))}
    \label{12},
\end{equation}
and to determine the Harrison transformed function $a$ in the next 
section,
\begin{equation}
    \mathcal{S}_{21}=-\frac{\mathcal{E}\bar{\mathcal{E}}}{2if}
    D_{\infty^{+}}\ln\frac{\Theta(u+\omega(\infty^{+}))
    }{\Theta(u+\omega(\infty^{+})+\omega(\bar{\xi}))}
    \label{21}.
\end{equation}
Using a degenerated version of Fay's trisecant identity \cite{fay} (see 
\cite{prd2} for the present case), we can write the above relations 
free of derivatives,
\begin{eqnarray}
    S_{12}&=&\frac{\rho}{2f}\left(\frac{\Theta[m](u)\Theta[m](u+
    2\omega(\infty^{-})+\omega(\bar{\xi}))}{L\Theta[m](u+
    \omega(\infty^{-})+\omega(\bar{\xi}))\Theta[m](u+
    \omega(\infty^{-}))}-1\right)\nonumber\\
    S_{21}&=&-\frac{\rho\mathcal{E}\bar{\mathcal{E}}}{2f}
    \left(\frac{\Theta[m](u)\Theta[m](u+
    2\omega(\infty^{+})+\omega(\bar{\xi}))}{L\Theta[m](u+
    \omega(\infty^{+})+\omega(\bar{\xi}))\Theta[m](u+
    \omega(\infty^{+}))}-1\right)
    \label{Z'},
\end{eqnarray}
where 
\begin{equation}
    L=\frac{\Theta(\omega(\infty^{-}))\Theta(\omega(\infty^{-})+
    \omega(\bar{\xi}))}{\Theta(0)\Theta(
    \omega(\bar{\xi}))}
    \label{C}.
\end{equation}

To construct the solution for the counter-rotating disks in \cite{prd3}, 
we used an algebraic approach which made it possible to establish 
algebraic relations between the metric functions at the disk. 
Let us recall that a divisor $X$ on $\Sigma_{g}$
is a formal symbol  $X=n_{1}P_{1}+\ldots+ n_{k}P_{k}$ with $P_{i}\in 
\Sigma$ and $n_{i}\in \mbox{Z}$. The degree of a divisor is 
$\sum_{i=1}^{k}n_{i}$.  The Riemann vector $K_{R}$ is 
defined by the condition that $\Theta(\omega(W)+K_{R})=0$ if $W$ is a 
divisor of degree $g-1$ or less. We use here and in the following the 
notation $\omega(W)=\int_{P_{0}}^{W}\mathrm{d}\omega
=\sum_{i=1}^{g-1}\omega(W_{i})$. 
Note that  the Riemann vector can be 
expressed through half-periods in the case of a hyperelliptic surface.
We define the divisor $X=\sum_{i=1}^{g}X_{i}$ 
as the solution of the Jacobi inversion problem ($i=1,\ldots,g$)
\begin{equation}
	\omega(X)-\omega(D)=u
	\label{eq18},
\end{equation}
where the divisor $D=\sum_{i=1}^{g}E_{i}$ (this corresponds to a 
choice of the characteristic $[m]$ in (\ref{rel1a})). 
With the help of these divisors, we can write (\ref{rel1a}) in the form
\begin{equation}
	\ln \mathcal{E}=\int_{D}^{X}\frac{\tau^{g}d\tau}{
	\mu(\tau)}-\frac{1}{2\pi \mathrm{i}}
	\int_{\Gamma}^{}\ln G \frac{\tau^{g}d\tau}{\mu(\tau)}
	\label{eq19},
\end{equation}
Additional information follows from the reality of $u$ 
which leads to $\omega(X)-\omega(D)=\omega(\bar{X})-\omega(\bar{D})$. 
The reality condition for $X$ implies via Abel's theorem the existence 
of a meromorphic function $R$ with poles in $\bar{X}+D$ and zeros in 
$X+\bar{D}$ (which is a rational function in the fundamental polygon),
\begin{equation}
    R(K)=const \frac{\prod_{i=1}^{g}(K-E_{i})(K-\bar{\mathcal{E}}_{i})-Q_{0}(K)\mu(K)}{
    \prod_{i=1}^{g}(K-\bar{X}_{i})(K-E_{i})}
    \label{abel},
\end{equation}
where $Q_{0}(K)=x_{0}+x_{1}K+\ldots+xK^{g-1}$ is a polynomial in 
$K$ with purely imaginary coefficients and $x=ibe^{-2U}$. 
The coefficients $x_{i}$ are related 
to $X$ via the relation 
\begin{equation}
    (1-x^{2})\prod_{i=1}^{g}(K-X_{i})(K-\bar{X}_{i})=
    \prod_{i=1}^{g}(K-E_{i})(K-\bar{\mathcal{E}}_{i})-Q_{0}^{2}(K)(K-\xi)(K-\bar{\xi})
    \label{abela}.
\end{equation}
We can use the existence of the 
rational function $R$ to calculate 
certain integrals of the third kind as 
\begin{equation}
    \frac{\Theta(u+\omega(P))\Theta(u+\omega(P^{\sigma})+
\omega(\bar{\xi})) 
}{\Theta(u+\omega(P^{\sigma}))\Theta(u+\omega(P)+\omega(\bar{\xi}))}
=\exp\left(\int_{X+\bar{D}}^{\bar{X}+D}d\omega_{P^{\sigma}P}
\right)=\frac{R(P)}{R(P^{\sigma})}
    \label{abel2}.
\end{equation}
This makes it possible to give an algebraic expression for $S_{12}$ 
and $S_{21}$. We can write $S_{12}$ in the form
\begin{equation}
    S_{12}=\frac{1}{i(\mathcal{E}+\bar{\mathcal{E}})}
    D_{\infty^{+}}\left(\int_{X}^{\bar{X}}d\omega_{PQ}
    \right)=\frac{1}{i(\mathcal{E}+\bar{\mathcal{E}})}
    D_{\infty^{+}}\left(\int_{X+\bar{D}}^{\bar{X}+D}d\omega_{PQ}
    +\int_{D}^{\bar{D}}d\omega_{PQ}
    \right)
    \label{12a}
\end{equation}
with $Q$ independent of $P$. 
The second integral can be reexpressed in terms of theta functions and 
be calculated with the help of so-called root 
functions:
The quotient of two theta functions with the same argument but 
different characteristic is a  root function which means that 
its square is a function on $\Sigma_{g}$. 
Let  $P_{i}$, $i=1,\ldots, 2g+2$, be 
the branch points of a hyperelliptic Riemann surface $\Sigma_{g}$ of genus $g$ 
and $A_{j}=\omega(P_{j})$ with $\omega(P_{1})=0$. Furthermore let 
$\{i_{1},\ldots, i_{g}\}$ and $\{j_{1},\ldots,j_{g}\}$ be two sets of 
numbers in $\{1,2,\ldots,2g+2\}$. Then the following 
equality holds for an 
arbitrary point $P\in \Sigma_{g}$,
\begin{equation}
	\frac{\Theta\left[K_{R}+\sum_{k=1}^{g}A_{i_{k}}\right]\left(
	\omega(P)\right)}{\Theta\left[K_{R}+
	\sum_{k=1}^{g}A_{j_{k}}\right]\left(
	\omega(P)\right)}=c_{1}\sqrt{\frac{(K-E_{i_{1}})\ldots(K-E_{i_{g}})}{
	(K-E_{j_{1}})\ldots(K-E_{j_{g}})}}
	\label{root1},
\end{equation}
where $c_{1}$ is a constant independent of $K$.

This implies 
\begin{equation}
    \exp(\int_{D}^{\bar{D}}d\omega_{P^{\sigma}\infty^{+}})
= \pm \prod_{i=1}^{g}\sqrt{\frac{K-F_{i}}{K-E_{i}}}
    \label{root2}.
\end{equation}
Thus we get with (\ref{abel2})
\begin{equation}
    \mathcal{S}_{12}=\frac{1}{i(\mathcal{E}+\bar{\mathcal{E}})}\left(\frac{1}{2}
    \sum_{i=1}^{g}(X_{i}-\bar{X}_{i})+\frac{1}{1-x^{2}}
    \left(\frac{x}{2}\left(\sum_{i=1}^{g}(E_{i}+\bar{E}_{i})-(\xi+\bar{\xi})
    \right)+x_{g-2}\right)\right),
    \label{12b}
\end{equation}
and similarly
\begin{equation}
    \mathcal{S}_{21}=-\frac{\mathcal{E}\bar{\mathcal{E}}}{i(\mathcal{E}+
    \bar{\mathcal{E}})}\left(\frac{1}{2}
    \sum_{i=1}^{g}(X_{i}-\bar{X}_{i})-\frac{1}{1-x^{2}}
    \left(\frac{x}{2}\left(\sum_{i=1}^{g}(E_{i}+\bar{E}_{i})-(\xi+\bar{\xi})
    \right)+x_{g-2}\right)\right).
    \label{21b}
\end{equation}

\subsection{Counter-rotating dust disk}
We summarize results of \cite{prd3} and \cite{prd4} on
disks which can be interpreted as two
counter-rotating components of pressureless matter, so-called dust. 
The surface energy-momentum tensor $S^{\alpha\beta}$ 
of these models, where $\alpha$ and $\beta$ 
stand for the $t$, $\rho$ and $\phi$ components, 
is defined on the
hypersurface $\zeta=0$.  The tensor $S^{\alpha\beta}$ is related to the 
energy-momentum tensor $T^{\alpha\beta}$ which appears in the Einstein 
equations $G^{\mu\nu}=8\pi T^{\mu\nu}$ (we use units in which the 
Newtonian gravitational constant, the dielectric constant 
and the velocity of light are equal 
to 1) via $T^{\alpha\beta}=S^{\alpha\beta}e^{k-U}\delta(\zeta)$. 
The tensor $S^{\alpha\beta}$ can be written in the form
\begin{equation} 
    S^{\alpha\beta}=\sigma_+ u^{\alpha}_+ u^{\beta}_+ +\sigma_{-}
u^{\alpha}_- u^{\beta}_- \label{2.0},
\end{equation}
where $u_{\pm}=(1,0,\pm \Omega)$.  We gave an explicit solution
for disks with constant angular velocity $\Omega$ and constant
relative density
$\gamma=(\sigma_{+}-\sigma_{-})/(\sigma_{+}+\sigma_{-})$.  This class
of solutions is characterized by two real parameters $\lambda$ and
$\delta$ which are related to $\Omega$ and $\gamma$ and the metric
potential $U_{0}=U(0,0)$ at the center of the disk via
\begin{equation}
\lambda=2\Omega^{2}e^{-2U_{0}} \label{2.1}
\end{equation}
and
\begin{equation}
\delta=\frac{1-\gamma^{2}}{\Omega^{2}} \label{2.2}.
\end{equation} 
We put the radius $\rho_{0}$ of the disk equal to 1 unless otherwise
noted.  Since the radius appears only in the combinations
$\rho/\rho_{0}$, $\zeta/\rho_{0}$ and $\Omega \rho_{0}$ in the physical
quantities, it does not have an independent role.  It is always
possible to use it as a natural length scale unless
it tends to 0 as in the case of the ultrarelativistic limit of the one
component disk.  The Ernst potential
will be discussed in dependence of
the parameters $\epsilon=z_{R}/(1+z_{R})=1-e^{U_{0}}$  and $\gamma$, 
where $z_{R}$ is the redshift of photons emitted at the center of the 
disk and detected at infinity.

The solution is given on a surface of genus 2 where
the branch points of the Riemann surface are given by the relation 
$E_{1}=-\bar{E}_{2}$ and
$E:=E_{1}^{2}=\alpha+\mathrm{i}\beta$ with $\alpha$, $\beta$ 
given by
\begin{equation} \alpha=-1+\frac{\delta}{2},\quad
\beta=\sqrt{\frac{1}{\lambda^{2}}+\delta -\frac{\delta^{2}}{4}}
\label{2.3}.
\end{equation}
The function $G$ in (\ref{rel1a}) reads
\begin{equation}
G(\tau)=\frac{\sqrt{(\tau^{2}-\alpha)^2+\beta^2}+\tau^{2}+1}{
\sqrt{(\tau^{2}-\alpha)^2+\beta^2}-(\tau^{2}+1)}.  \label{2.4}
\end{equation}
We note that with $\alpha$ and $\beta$ given, the
Riemann surface is completely determined at a given point in the
spacetime, i.e.\ for a given value of $\xi$. 

Regularity of the solutions in the exterior of the disk
restricts the physical parameters to
$0\leq \delta\leq
\delta_{s}(\lambda):=2\left(1+\sqrt{1+1/\lambda^{2}}\right)$ and $0<\lambda
\leq \lambda_{c}$ where $\lambda_{c}(\gamma)$ is the smallest value of
$\lambda$ for which $\epsilon=1$.  
The range of the physical parameters is restricted by the following 
limiting cases:\\
Newtonian limit: $\epsilon=0$ ($\lambda=0$), i.e.\ small 
velocities $\Omega \rho_{0}$ and small redshifts in the disk. 
The function $e^{2U}$ tends independently of $\gamma$  
to $1+\lambda U_{N}$, where $U_{N}$ is the Maclaurin disk 
solution, and $b$ is of order 
$\Omega^{3}$. \\
Ultrarelativistic limit: $\epsilon=1$, i.e.\ 
diverging central redshift. For $\gamma\neq1$ it is reached for 
$\lambda_{c}=\infty$. The solution describes a disk of finite 
extension with diverging 
central redshift.
For $\gamma=1$, the limit is reached for  $\lambda_{c}=4.629\ldots$. 
In this case the solution has a singular axis and is not 
asymptotically flat. This behavior can be interpreted as the limit 
of a vanishing disk radius. With this rescaling the solution in the 
exterior of the disk can be interpreted as the extreme Kerr solution 
(see \cite{prd4} and references given therein).
\\
Static limit: $\gamma=0$ ($\delta=\delta_{s}(\lambda)$). 
In this limit,  the solution belongs to the Morgan and Morgan class 
\cite{morgan}. 
\\
One component: $\gamma=1$ ($\delta=0$), i.e.\ no counter-rotating 
matter in the disk. This is the disk of \cite{bawa}, 
\cite{neugebauermeinel}.

Analytic formulas for the complete metric in terms of theta functions 
are given in \cite{prd4}. To evaluate the hyperelliptic integrals in 
the expressions for the metric we use the numerical methods and the 
cut-system of \cite{prd4}.

At the disk the branch points $\xi,\bar{\xi}$ lie on the contour $\Gamma$ 
which 
implies that care has to be taken in the evaluation of the line integrals. 
The situation is however simplified by the equatorial symmetry of 
the solution which is reflected by the additional involution $K\to -K$
of the Riemann surface $\Sigma_{2}$ for $\zeta=0$. 
This makes it possible to perform the reduction $K^{2}\to \tau$ and 
to express the metric 
in terms of elliptic theta functions (see \cite{prd2}). 
We denote with $\Sigma_{w}$ the elliptic Riemann surface defined by
$\mu_{w}^{2}=(\tau+\rho^{2})((\tau-\alpha)^{2}+\beta^{2})$, and let $dw$
be the associated differential of the first kind with
$u_{w}=\frac{1}{i\pi}\int_{-\rho^{2}}^{-1}\ln G(\sqrt{\tau})dw(\tau)$. 
 We cut the 
surface in a way that the $a$-cut is a closed contour in the upper 
sheet around the cut $[-\rho^{2},\bar{E}]$ and that the $b$-cut starts 
at the cut $[\infty,E]$. The Abel map $w$ is defined for $P\in 
\Sigma_{w}$ as $w(P)=\int_{\infty}^{P}dw$.
Then the real part of the Ernst potential at the disk can be written as
\begin{eqnarray}
e^{2U}&=&\frac{1}{Y-\delta}\left(-\frac{1}{\lambda}-\frac{Y}{\delta}
\left(\frac{\frac{1}{\lambda^{2}}+\delta}{
\sqrt{\frac{1}{\lambda^{2}}+\delta\rho^{2}}}-\frac{1}{\lambda}\right)
\right.  \nonumber\\ &&\left. 
+\sqrt{\frac{Y^{2}((\rho^{2}+\alpha)^{2}+\beta^{2})}{
\frac{1}{\lambda^{2}}+\delta\rho^{2}}-2Y(\rho^{2}+\alpha) +
\frac{1}{\lambda^{2}}+\delta\rho^{2}}\right) \label{y1},
\end{eqnarray}
where
\begin{equation}
Y=\frac{\frac{1}{\lambda^{2}}+\delta\rho^{2}}{\sqrt{(\rho^{2}+\alpha)^{2}
+\beta^{2}}}\frac{\vartheta_{3}^{2}(u_{w})}{\vartheta_{1}^{2}(u_{w})}
\label{y2}.
\end{equation}
It was shown that there exist algebraic relations between the
real and imaginary parts of the Ernst potential, 
\begin{equation}
\frac{\delta^{2}}{2}(e^{4U}+b^{2})= \left(\frac{1}{\lambda}-\delta
e^{2U}\right)\left(\frac{\frac{1}{\lambda^{2}
}+\delta}{\sqrt{\frac{1}{\lambda^{2}}+\delta \rho^{2}}}-
\frac{1}{\lambda}\right)+\delta\left(\frac{\delta+\rho^{2}}{2}-1\right)
\label{2.9},
\end{equation}
and the function $Z:=(a-a_{0})e^{2U}$ 
\begin{equation}
   Z^{2}-\rho^{2}+\delta e^{4U}=\frac{2}{\lambda}e^{2U} \label{2.10a}.
\end{equation}
Moreover we have
\begin{equation}
    ix_{0}=-\frac{Z}{\delta f}\left(\frac{1/\lambda^{2}+
    \delta}{\sqrt{1/\lambda^{2}+\delta \rho^{2}}}-\frac{1}{\lambda}
    \right)
    \label{x0}.
\end{equation}

At the disk, the normal derivatives of the 
metric functions are discontinuous which leads via the Einstein 
equations to a $\delta$-type energy-momentum tensor. To determine 
this tensor, it 
seems best to use Israel's invariant 
junction conditions for matching spacetimes across non-null hypersurfaces 
\cite{israel}. As in \cite{prd3} we
get
\begin{eqnarray}
	s_{1}:=- 4\pi s_{00} & = & \left(k_{\zeta}-2U_{\zeta}
	\right)e^{2U},
	\nonumber \\
    s_{2}:=- 4\pi ( s_{03}-as_{00}) & = & -\frac{1}{2}a_{\zeta}e^{2U}
	\nonumber ,\\
	s_{3}:=-4\pi (s_{33}-2a s_{03}+a^2 s_{00})& = & -k_{\zeta}\rho^2e^{-2U}
	\label{16.7},
\end{eqnarray}
where we define the tensor $s^{\mu\nu}$ via 
$T^{\mu\nu}=s^{\mu\nu}e^{2(k-U)}\delta(\zeta)$.
The right-hand sides in (\ref{16.7}) can be expressed in terms of 
$\rho$-derivatives in the disk via
\begin{eqnarray}
    (e^{2U})_{\zeta} & = & \frac{Z^{2}+\rho^{2}+\delta e^{4U}}{2Z\rho}b_{\rho}
    \nonumber  \\
    b_{\zeta} & = & -\frac{Z^{2}+\rho^{2}+\delta e^{4U}}{2Z\rho}
    (e^{2U})_{\rho}+\frac{e^{2U}}{Z}
    \label{diskz}.
\end{eqnarray}
This makes it directly 
possible to determine all quantities in the disk in 
terms of elliptic functions.

\section{Einstein-Maxwell equations and Harrison transformations}
\subsection{Field equations and group structure}
To treat the Einstein-Maxwell equations in vacuum in the 
stationary axisymmetric case we use again the metric (\ref{vac1}).
We gauge the electromagnetic 4-potential in a way that it has only 
non-vanishing $A_{t}$ and $A_{\phi}$ components. It is subject to the 
Maxwell equations
\begin{equation}
    F_{\mu\nu}{}^{;\nu}=0, \quad ^{*}F_{\mu\nu}{}^{;\nu}=\frac{1}{2}
    \epsilon_{\mu\nu\alpha\beta}F^{\alpha\beta}{}^{;\nu}=0, \quad
    F_{\mu\nu}=A_{\mu,\nu}-A_{\nu,\mu}
    \label{max}.
\end{equation}
The 
Einstein equations lead to 
\begin{equation}
    R_{\mu\nu}=F_{\mu\lambda}F^{\lambda}_{\nu}-\frac{1}{4}g_{\mu\nu}
    F_{\kappa\lambda}F^{\kappa\lambda}
    \label{einstein}.
\end{equation}
The vacuum Einstein-Maxwell equations in the stationary axisymmetric 
case are equivalent to the complex Ernst equations \cite{ernst2}
\begin{eqnarray}
    f\Delta \mathcal{E} & = & (\nabla \mathcal{E}+2\bar{\Phi} \nabla 
    \Phi)\nabla \mathcal{E},
    \nonumber  \\
    f\Delta \Phi & = & (\nabla \mathcal{E}+2\bar{\Phi} \nabla 
    \Phi)\nabla \Phi,
    \label{ernst}
\end{eqnarray}
where $\Delta$ and $\nabla$ are the standard differential operators in 
cylindrical coordinates, where the potentials $\mathcal{E}$ and $\Phi$ 
are independent of $\phi$, and 
where $\Re \mathcal{E}=f-\Phi \bar{\Phi}$ and $f=e^{2U}$. 
The remaining metric functions follow via 
\begin{eqnarray}
    k_{\xi} & = & \frac{\xi-\bar{\xi}}{f}\left( \frac{1}{4f}(
    \mathcal{E}_{\xi} 
    +2\bar{\Phi}\Phi_{\xi} )(\bar{\mathcal{E}}_{\xi} 
    +2\Phi\bar{\Phi}_{\xi} ) - \Phi_{\xi}\bar{\Phi}_{\xi}\right)
    \label{k}  \\
    a_{\xi} & = & \frac{\rho}{f^{2}}\left(\mathrm{i}(\Im 
    \mathcal{E})_{\xi} +\Phi\bar{\Phi}_{\xi}-\bar{\Phi}\Phi_{\xi}\right)
    \label{a}
\end{eqnarray}
from $\mathcal{E}$ and $\Phi$. The electromagnetic potentials are 
related to $\Phi$ via $A_{t}=\Re \Phi$ and 
\begin{equation}
(\Im \Phi)_{\xi}=\frac{i}{\rho}f
(A_{\phi,\xi}-aA_{t,\xi})
    \label{magnetic}.
\end{equation}
For $\Phi=0$ one obtains the pure Einstein case of the previous 
section.

The system (\ref{ernst}) is again integrable and can be treated as 
the integrability condition of a linear differential system as in the 
pure Einstein case (see \cite{meudon}). 
Even without axial symmetry 
the stationary Einstein-Maxwell equations are equivalent 
to a $SU(2,1)/S[U(1,1)\times U(1)]$ nonlinear sigma model. 
This group 
can be parametrized by trigonal $3\times3$ matrices $V$,
\begin{equation}
    V=\left(
    \begin{array}{ccc}
        \sqrt{f} & 0 & 0  \\
        i\sqrt{2}\Phi & 1 & 0  \\
        (b+i|\Phi|^{2})/\sqrt{f} & 
        (\sqrt{2}\bar{\Phi})/\sqrt{f} & 1/\sqrt{f}
    \end{array}\right)
    \label{group1}.
\end{equation}
The matrix $V$ satisfies 
\begin{equation}
    V^{\dagger}\eta V=\eta , \quad \eta =\left(
    \begin{array}{ccc}
        0 & 0 & i  \\
        0 & 1 & 0  \\
        -i & 0 & 0
    \end{array}\right)
    \label{eta},
\end{equation}
i.e.\ it is unitary with respect to the metric $\eta$ of $SU(2,1)$. 
The action of $\mathcal{G}\in SU(2,1)$ on $V$ is $V\to 
h(V,\mathcal{G})V\mathcal{G}^{-1}$ where 
$h(V,\mathcal{G})\in S[U(1,1)\times U(1)]$ restores the triangular gauge of $V$. 
To obtain a gauge invariant parametrization, one introduces $\chi:= 
\Xi V^{\dagger}\Xi V$ with $\Xi=\mbox{diag}(1,-1,1)$ 
on which the action of $\mathcal{G}\in SU(2,1)$ is given by $\chi 
\to \Xi(\mathcal{G}^{-1})^{\dagger}\Xi\chi \mathcal{G}^{-1}$. We have
\begin{equation}
    \chi=\left(
    \begin{array}{ccc}
        f-2|\Phi|^{2}+(b^{2}+|\Phi|^{4})/f & 
        \sqrt{2}\bar{\Phi} 
        (b-i|\Phi|^{2}+if)/f & (b-i|\Phi|^{2})/f  \\
        -\sqrt{2}\Phi(b+i|\Phi|^{2}-if)/f & 
        1-2|\Phi|^{2}/f & -\sqrt{2}\Phi/f  \\
        (b+i|\Phi|^{2})/f & \sqrt{2}\bar{\Phi}/f & 
        1/f
    \end{array}\right)
    \label{group2}.
\end{equation}
The $SU(2,1)$ symmetry can be used to generate solutions by the action of an 
element $\mathcal{G}$. We list the infinitesimal transformations  and their 
consequences:
$\left(
\begin{array}{ccc}
    0 & 0 & 0  \\
    \theta_{1} & 0 & 0  \\
    \theta_{2} & \theta_{3} & 0
\end{array}\right)$ are gauge transformations which add physically 
irrelevant constants to $\Im \mathcal{E}$ and $\Im \Phi$, $\left(
\begin{array}{ccc}
    0 & 0 & \theta  \\
    0 & 0 & 0  \\
    0 & 0 & 0
\end{array}\right)$ is an Ehlers transformation 
which changes $f\to b$, $\left(
\begin{array}{ccc}
    i\theta & 0 & 0  \\
    0 & -2i\theta & 0  \\
    0 & 0 & i\theta
\end{array}\right)$ is an electromagnetic duality transformation, 
$\left(
\begin{array}{ccc}
    \theta & 0 & 0  \\
    0 & 0 & 0  \\
    0 & 0 & -\theta
\end{array}\right)$ is a scale transformation, $f,b,\Phi\to 
e^{\theta}f, e^{\theta}b, e^{\theta/2}\Phi$, and 
$\left(
\begin{array}{ccc}
    0 & -i\theta & 0  \\
    0 & 0 & \bar{\theta}  \\
    0 & 0 & 0
\end{array}\right)$ is a Harrison transformation \cite{harrison} 
which changes $f\to \Phi$.

\subsection{Harrison transformations}
Here we are interested to generate charged solutions from non-charged 
ones via a Harrison transformation, i.e.\ to generate solutions to the 
Einstein-Maxwell equations containing one additional constant 
parameter which is related to the charge. 
We assume that the original solutions are  equatorially symmetric 
and asymptotically flat, i.e.\
$f\to 1$, $\Phi\to0$ and $b\to0$ for $|z|\to 
\infty$. To ensure that the transformed solutions have the same 
asymptotic behavior, one has to use a scale transformation 
together with 
a transformation which changes $\Phi$ and $b$ by some constant. By 
exponentiating the matrices of the $SU(2,1)$ transformations, we 
consider a transformation of the form 
\begin{equation}
    \mathcal{G}=\left(
    \begin{array}{ccc}
	1 & i\theta_{1} & -i\theta_{1}\bar{\theta_{1}}/2  \\
        0 & 1 & -\bar{\theta_{1}}  \\
        0 & 0 & 1
    \end{array}\right)\mbox{diag}(e^{-\theta_{2}},1,e^{\theta_{2}})
    \left(
    \begin{array}{ccc}
        1 & 0 & 0  \\
        -\theta_{3} & 1 & 0  \\
        -\theta_{4}+\theta_{3}\theta_5/2 & -\theta_5 & 1
    \end{array}\right).
    \label{group6}
\end{equation}
Since the asymptotic 
conditions imply that $\chi$ and $\chi'$ are  the 
unit matrix at infinity, the matrix $\mathcal{G}$ must satisfy the condition 
 $\Xi \mathcal{G}^{\dagger}\Xi =\mathcal{G}^{-1}$. 
This leads with (\ref{group6}) to
\begin{equation}
    e^{\theta_2}=\frac{1}{1-\theta_1\bar{\theta_1}/2},\quad 
    \theta_3=i\bar{\theta_1}, \quad \theta_4=0,\quad \theta_5=\theta_1
    \label{group9}.
\end{equation}
The matrix $\mathcal{G}$ thus takes the form
\begin{equation}
    \mathcal{G}=\frac{1}{1-\theta_1\bar{\theta_1}/2}\left(
    \begin{array}{ccc}
        1 & i\theta_1 & -i\theta_1\bar{\theta_1}/2  \\
        -i\bar{\theta_1} & 1+\theta_1\bar{\theta_1}/2 & -\bar{\theta_1}  \\
         i\theta_1\bar{\theta_1}/2& -\theta_1 & 1
    \end{array}
    \right)
    \label{group10}.
\end{equation}
If we transform an Ernst potential in the case $\Phi=0$, we end up 
with 
\begin{equation}
    \Phi'=-\frac{\theta_1}{\sqrt{2}}\frac{\theta_1\bar{\theta_1}(f^{2}+b^{2})
    /2-(1+\theta_1\bar{\theta_1}/2)f+1-ib(1-\theta_1\bar{\theta_1}/2)}{(1-\theta_1\bar{\theta_1}f/2 
    )^{2}+(\theta_1\bar{\theta_1}/2)^{2}b^{2}}
    \label{group11}.
\end{equation}
We are interested in transformations which preserve the equatorial
symmetry, i.e.\ $f(-\zeta)=f(\zeta)$, $b(-\zeta)=-b(\zeta)$ and 
$\Phi(-\zeta)=\bar{\Phi}(\zeta)$. This implies for (\ref{group11}) 
that $\theta_1$ must be real which rules out magnetic monopoles. We 
put $q=\theta_1/\sqrt{2}$ and sum up the results for the transformed 
potentials:
\begin{eqnarray}
    f' & = & \frac{(1-q^{2})^{2}f}{(1-q^{2}f)^{2}+q^{4}b^{2}}
    \label{f'}  \\
    b' & = & \frac{(1-q^{4})b}{(1-q^{2}f)^{2}+q^{4}b^{2}}
    \label{b'}  \\
    \Phi' & = & -q\frac{(1-f)(1-q^{2}f)+q^{2}b^{2} +ib(1-q^{2})}{
    (1-q^{2}f)^{2}+q^{4}b^{2}}
    \label{Phi'}.
\end{eqnarray}
The real parameter $q$ has to be in the region $0<|q|<1$, for $q>1$ 
the transformed spacetime would have a negative mass. The value 
$q=0$ corresponds to the untransformed solution. The above formulas 
imply that the functions $f'$, $b'$ and $\Phi'$ are analytic 
where the original functions are analytic.

The metric function $k$ is invariant under the action of $SU(2,1)$ 
transformations. 
To determine the transformed metric function $a$, we 
consider the matrix $S$ in (\ref{12}). If we go over from 
$2\times 2$ matrices to 
$3\times 3$ matrices according to the rule
\begin{equation}
    \left(
    \begin{array}{cc}
        A_{11} &A_{12}   \\
        A_{21} & A_{22}
    \end{array}\right)\to  \left(
    \begin{array}{ccc}
        A_{11} & 0 & A_{12}  \\
	0 & 1 & 0 \\
        A_{21} & 0 & A_{22}
    \end{array}\right)
    \label{rule},
\end{equation}
the matrix $\mathcal{G}$ acts on  $S$ as on $\chi$.
Thus we get with (\ref{group10}) 
\begin{equation}
    \mathcal{S}_{12}'=\frac{1}{(1-q^{2})^{2}}(\mathcal{S}_{12}+
    2iq^{2}-q^{4}\mathcal{S}_{21})
    \label{ehlers4},
\end{equation}
which is in accordance with (\ref{a}). This implies with (\ref{a4})
for the function $a'$
\begin{equation}
    a'-a_{0}'=-\frac{2}{(1-q^{2})^{2}}(\mathcal{S}_{12}-q^{4}
    \mathcal{S}_{21})
    \label{a'}.
\end{equation}
To determine $a_{0}'$, one has to consider $\mathcal{S}_{12}$ and $
\mathcal{S}_{21}$ on 
the axis. In the limit $\rho\to0$, there is a non-trivial contribution 
from the quotient of theta functions in (\ref{Z'}) which diverges as 
$1/\rho$. Repeating the considerations of \cite{prd2} in the 
calculation of $a_{0}$, one finds that the axis potentials can be 
expressed in terms of theta functions on the surface $\tilde{\Sigma}$ 
of genus $g-1$ given by $\tilde{\mu}^{2}=\prod_{i=1}^{g}(K-E_{i})(K-
\bar{E}_{i})$. Denoting quantities on $\tilde{\Sigma}$ with a tilde, 
one has with \cite{prd2} on the axis
\begin{equation}
    \frac{\mathcal{S}_{21}}{\mathcal{S}_{12}}=
    \frac{\tilde{\Theta}[\tilde{m}](\tilde{u}+
        2\tilde{\omega}(\infty^{-}))}{\tilde{\Theta}[\tilde{m}](\tilde{u}+
        2\tilde{\omega}(\infty^{+}))}
    \label{a0a}.
\end{equation}
In the equatorially symmetric case, this quotient is identical to one 
since $2\tilde{\omega}(\infty^{+})$ is a half period on 
$\tilde{\Sigma}$ (see \cite{prd2}). Thus we have in this case 
\begin{equation}
    a_{0}'=a_{0}\frac{1+q^{4}}{(1-q^{2})^{2}}
    \label{Z0}.
\end{equation}

A well-known example is the Harrison
transformation of the Schwarzschild 
solution which leads to the Reissner-Nordstr\"om solution. There we 
have with  $r_{\pm}=\sqrt{(\zeta\pm m)^{2}+\rho^{2}}$ the relations
$f=(r_{+}+r_{-}-2m)/(r_{+}+r_{-}+2m)$, $b=0$ which implies with 
(\ref{Phi'}) 
\begin{equation}
    f'  =  \frac{(r_{+}+r_{-})^{2}-4m^{2}}{(r_{+}+r_{-}+2m')^{2}},
   \quad
    \Phi'  =  \frac{2Q}{r_{+}+r_{-}+2m'}
    \label{reissner},
\end{equation}
and $b'=0$. We thus get a static 
solution with mass $m'$ and charge $Q$ which are subject to the 
relation $m'{}^{2}-Q^{2}=m^{2}$.

\subsection{Asymptotic behavior}
We assume that the asymptotic behavior of the original solution, 
which is read off on the axis, is of the form 
$f=1-2M/|\zeta|$, $b=-2J/\zeta^{2}$ and $\Phi=Q/|\zeta|-iJ_{M}/\zeta^{2}$ plus 
terms of lower order in $1/|\zeta|$ where 
$M$ is the Arnowitt-Deser-Misner mass, $J$ the angular momentum, $Q$ 
the electric charge and $J_{M}$ the magnetic moment.  The same will 
hold for the Harrison transformed potentials. We find 
\begin{equation}
    M'=M\frac{1+q^{2}}{1-q^{2}}-\frac{2q}{1-q^{2}}Q,\quad 
    J'=J\frac{1+q^{2}}{1-q^{2}}-\frac{2q}{1-q^{2}}J_{M},
    \label{group12}
\end{equation}
and
\begin{equation}
    Q'=Q\frac{1+q^{2}}{1-q^{2}}-\frac{2q}{1-q^{2}}M,\quad 
    J_{M}'=J_{M}\frac{1+q^{2}}{1-q^{2}}-\frac{2q}{1-q^{2}}J
    \label{group12a}.
\end{equation}
It is interesting to note that the quantities $M^{2}-Q^{2}$ and $J^{2}
-J_{M}^{2}$ are invariants of the transformation. They are related to 
the Casimir operator of the $SU(2,1)$-group.
If the original solution was uncharged, 
the extreme relation $M'=\pm Q'$ is only possible 
in the limit $M=0$. 

A further invariant is the combination $J_{M}M-JQ$ which is of 
importance in relation to the gyromagnetic ratio 
\begin{equation}
    g_{M}=\frac{2MJ_{M}}{JQ}.
    \label{gyro}
\end{equation}
Relation (\ref{group12}) implies that $g_{M}$ is 
equal to 2 if $Q=J_{M}=0$ and $q\neq 0$. Thus all 
solutions which can be generated via a Harrison transformation from 
solutions with vanishing electromagnetic fields as the Kerr-Newman 
family from Kerr
have a gyromagnetic ratio of 2. Due to the invariance of 
$J_{M}M-JQ$ under Harrison transformations, a gyromagnetic 
ratio of 2 is not changed under the transformation.

\section{Charged disks}
In this section we will use the Harrison transformation 
(\ref{group10}) on the dust disk 
\cite{prl2} to generate a disk 
solution to the Einstein-Maxwell equation. We discuss the metric, 
several interesting limiting cases and the energy-momentum tensor.
\subsection{Metric functions}
Since the metric function $e^{2k}$ is invariant under Harrison 
transformations, we refer the reader to \cite{prd4} for a 
discussion. The metric function 
$f'$ in (\ref{f'}) is proportional to $f$, which implies that
the transformed solution vanishes exactly where the original solution 
has zeros. Since the set of 
zeros of $f$ just defines
the ergoregions, the transformed solution has the same ergoregions 
(if any) as the original solution. 

\begin{figure}[htb]
    \centering
     \epsfig{file=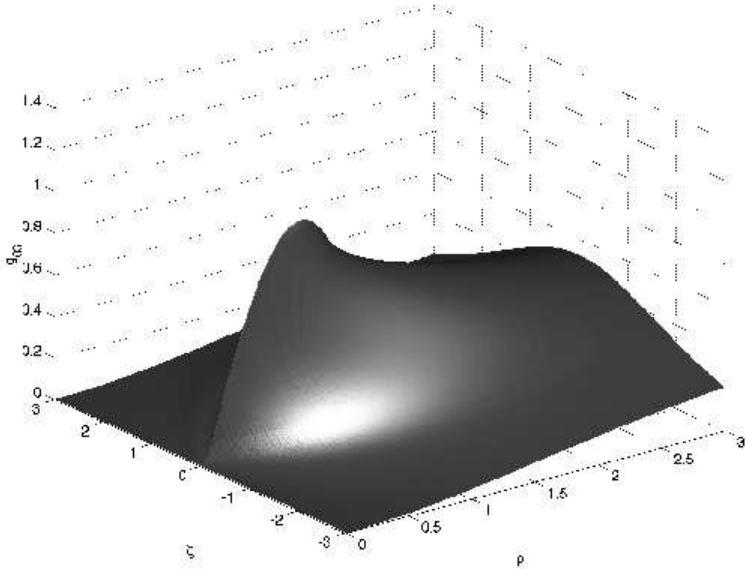,width=10cm}
    \caption{Metric function $g_{03}'$ for $\epsilon=0.85$, 
    $\gamma=0.95$ and $q=0.6$.}
    \label{Aprime}
\end{figure}

For small $q$, the functions $f$ and $b$ are essentially unchanged 
since they are quadratic in $q$. The electromagnetic potential $\Phi$ 
is in this limit with (\ref{Phi'}) of the form
\begin{equation}
    \Phi'=-q(1-f+ib)
    \label{qs}.
\end{equation}
For larger $q$, $|f'|$  becomes smaller near the origin. Since 
its asymptotic values are not changed, the growth  
rate towards infinity increases which is reflected by the mass 
formula (\ref{group12}). In the singular 
limit $q\to 1$, the function $f'$ is 
zero for all finite values of $z$, but one at infinity. The behavior 
for $b'$ is similar with the exception that $b'$ is odd and 
zero at infinity. 
The function $a$ also becomes singular in the limit $q\to 1$ 
which is reflected by the diverging factor $1/(1-q^{2})^{2}$ and the 
constant $a_{0}$ (\ref{Z0}) 
which just implies that one can no longer choose $a$ to be 
zero on the axis. In the metric function $g_{03}'=-a'f'$, the 
factors $(1-q^{2})^{2}$ just cancel and the function is only 
marginally changed with increasing $q$. The typical behavior of 
$g_{03}'$ for values of 
$q$ with $0<|q|<1$ can be seen in Fig. \ref{Aprime}.
The metric function is an even function in $\zeta$ which vanishes on 
the axis and at infinity. It is analytic except on the disk where the 
normal derivatives have a jump.

\begin{figure}[htb]
    \centering
     \epsfig{file=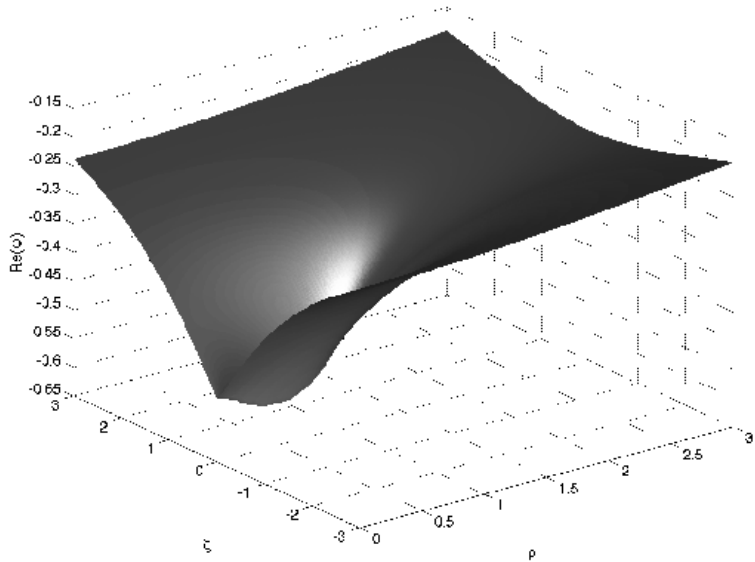,width=10cm}
    \caption{Real part of the electromagnetic potential $\Phi$ 
    for $\epsilon=0.85$, 
    $\gamma=0.95$ and $q=0.6$.}
    \label{RPhi'}
\end{figure}

\begin{figure}[htb]
    \centering
     \epsfig{file=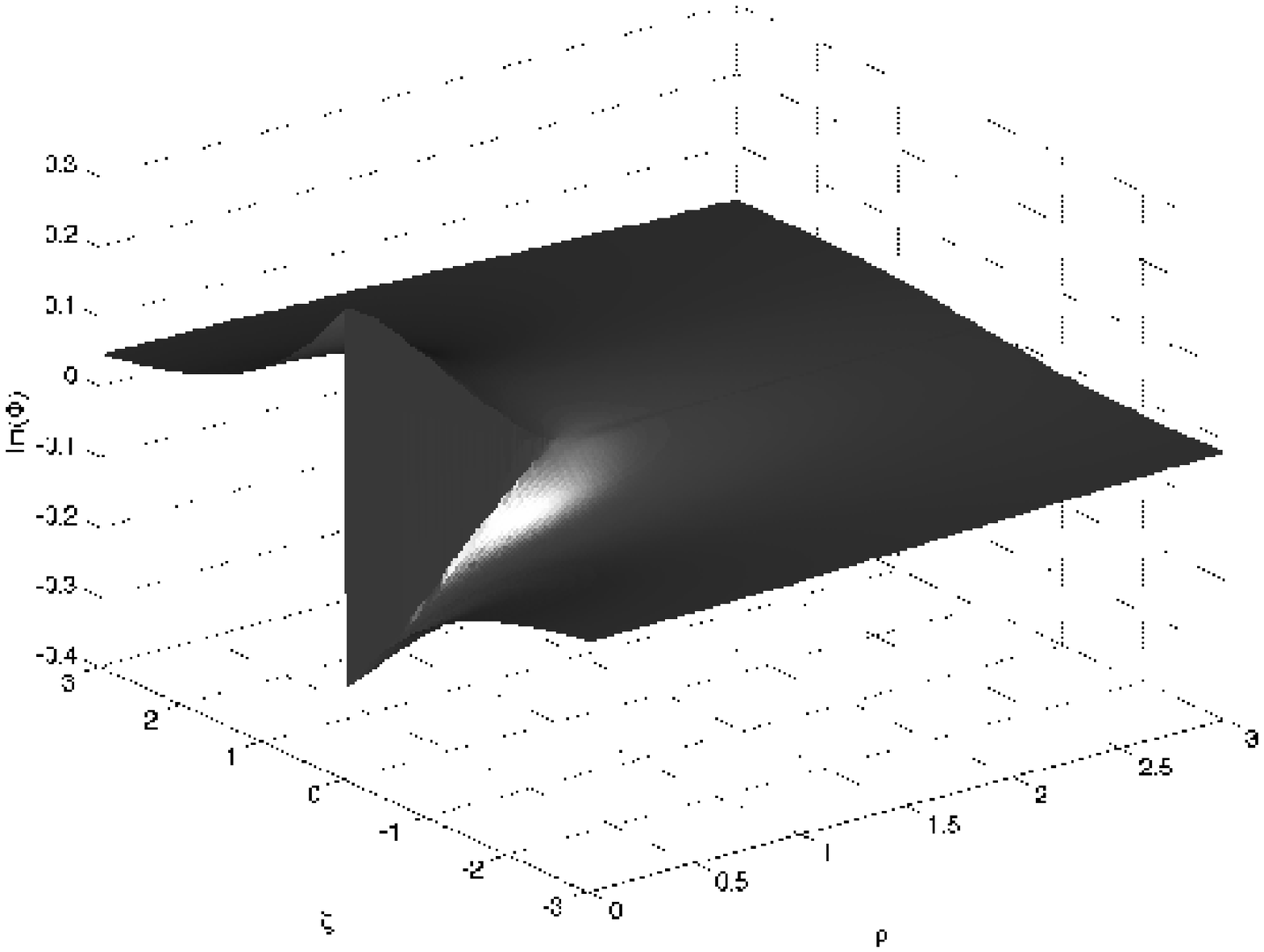,width=10cm}
    \caption{Imaginary part of the electromagnetic potential $\Phi$ 
    for $\epsilon=0.85$, 
    $\gamma=0.95$ and $q=0.6$.}
    \label{IPhi'}
\end{figure}

The electromagnetic potential tends to $-1$ in the limit $q\to 1$ for 
finite $\xi$, but is zero at infinity. The imaginary part is directly 
proportional to $b'$ as can be seen from (\ref{b'}) and (\ref{Phi'}).
We show a typical situation 
for values of $q$ with $0<|q|<1$ in Fig. \ref{RPhi'} for the real 
part and 
in Fig. \ref{IPhi'} for the imaginary part.
The real part of $\Phi$ is an even function in $\zeta$ which vanishes 
at infinity and has discontinuous normal derivatives at the disk.
The imaginary part of $\Phi$ is an odd function in $\zeta$ and has a 
jump at the disk. It vanishes at infinity.

Since $f'$ has the same zeros as $f$, the transformed solution has a 
diverging central redshift if the untransformed has, i.e.\ the 
ultrarelativistic limits coincide. In the case $\gamma\neq1$, one has 
a charged disk of finite extension 
with diverging central redshift. For $\gamma=1$, the 
solution in the exterior of the disk can be interpreted as 
an extreme Kerr-Newman metric which 
is obtained as a Harrison-transformed extreme Kerr metric. 

In the Newtonian limit $\lambda\to 0$, one has $f=1+\lambda U_{N}$ 
and $b=\lambda^{\frac{3}{2}}\tilde{b}$ in lowest order. This implies 
with (\ref{f'}) to (\ref{Phi'}) for $1-q^{2}>>\lambda$
\begin{equation}
    f'=1+\frac{1+q^{2}}{1-q^{2}}\lambda U_{N},\quad 
    b'=\frac{1+q^{2}}{1-q^{2}}\lambda^{\frac{3}{2}}\tilde{b}\quad
    \Phi'=-\frac{q}{1-q^{2}}(\lambda U_{N}+i
    \lambda^{\frac{3}{2}}\tilde{b}),
    \label{newt}
\end{equation}
The transformed solution thus has the same Newtonian and 
post-Newtonian behavior as the original metric and in addition an 
electromagnetic field. The magnetic field is of order $\Omega^{3}$ 
as $b'$.

Since the mass is of order $\lambda$ in the Newtonian limit, it is 
possible to have an extreme limit here with $M'=Q'$ as in the 
Reissner-Nordstr\"om solution. If we put $1-q^{2}=\kappa \lambda$ 
with $\kappa>0$, 
we get in the limit $\lambda\to0$ 
for (\ref{f'}) and (\ref{Phi'})
\begin{equation}
    f'=\frac{\kappa^{2}}{(\kappa-U_{N})^{2}}\quad \Phi'=
    \frac{U_{N}}{\kappa
    -U_{N}}
    \label{extreme2},
\end{equation}
a static solution similar to the extreme Reissner-Nordstr\"om 
solution, but with a jump in the normal derivatives of the metric 
functions at the disk and non-vanishing $f'$ at the 
origin.  
Since $U_{N}<0$ in the whole spacetime, the solution is regular in the 
exterior of the disk. Thus one gets a non-singular limit in the 
exterior of the disk for $q\to1$ 
in this case.

In the static limit one has $b'=\Im \Phi'=0$, since both are 
proportional to $b$,
\begin{equation}
    f'=\frac{(1-q^{2})f}{(1-q^{2}f)^{2}},\quad \Phi'=
    -q\frac{1-f}{1-q^{2}f}
    \label{static2}.
\end{equation}
The 
Harrison-transformed static solution is thus 
again static with vanishing 
magnetic field but non-zero electric field.

\subsection{Disk and energy-momentum tensor}
It remains to be checked whether and in which range of the 
parameters the energy-momentum tensor 
(\ref{tensor}) is physically acceptable. The above
discussion of the metric 
indicates the extreme behavior of the metric functions for $q$ close 
to one. It is plausible that the matter in the disk which is in the 
present example the
source of such an extreme metric  will in general not be  
physically acceptable. There can be maximal $q$ 
smaller than 1 for given $\lambda$ and $\delta$ which limits the 
physical range of the parameters. The matter in 
the disk will be interpreted by following Ledvinka et 
al.~\cite{zofka}.

\begin{figure}[htb]
    \centering
    \epsfig{file=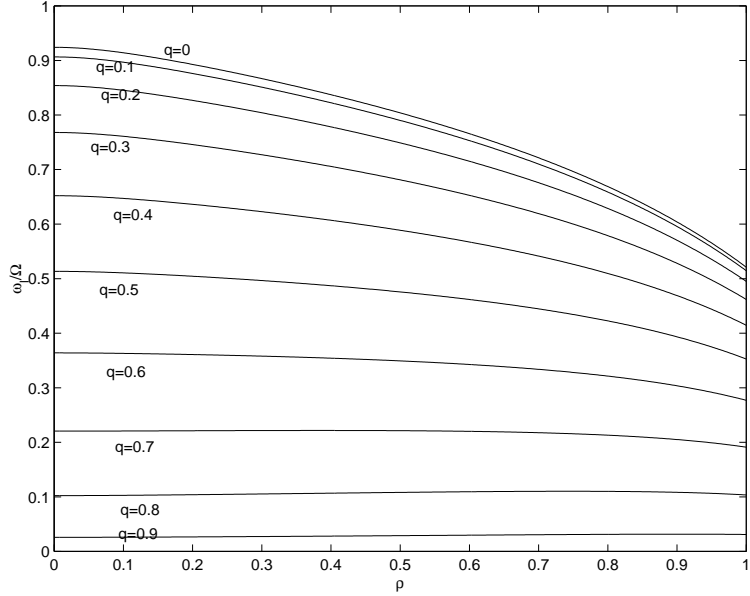,width=10cm}
    \caption{Angular velocity of the locally non-rotating observers 
    with respect to infinity for $\epsilon=0.85$, 
    $\gamma=0.95$, and $q=0,\ldots, 0.9$.}
    \label{ol}
\end{figure}

To discuss the energy-momentum tensor and the currents in the disk, 
it is helpful to use the algebraic relations (\ref{2.9}) to (\ref{x0})
between the metric functions which exist at the disk, and which imply 
similar relations between the transformed potentials. With 
(\ref{12b}) and (\ref{21b}), we get for $S_{21}$
\begin{equation}
    \mathcal{S}_{21}=\mathcal{E}\bar{\mathcal{E}}S_{12}+ix_{0}f
    \label{21d},
\end{equation}
and thus with (\ref{a'}) for the metric function $a'$
\begin{equation}
    (1-q^{2})^{2}(a-a_{0})'=(a-a_{0})
    \left(1+q^{4}\frac{2}{\delta}\left(\frac{1}{\delta\lambda}
    \left(\frac{1/\lambda^{2}+
    \delta}{\sqrt{1/\lambda^{2}+\delta \rho^{2}}}-\frac{1}{\lambda}
    \right)+\alpha+\frac{\rho^{2}}{2}\right)\right)
    \label{Z'3},
\end{equation}
where $a_{0}'$ is given by (\ref{Z0}). With this function we can 
determine the angular velocity $\omega_{l}$ with respect to infinity
of the locally non-rotating observers, 
for whom the metric is diagonal,
\begin{equation}
    \omega_{l}=-\frac{g_{03}}{g_{33}}=\frac{ae^{4U}}{\rho^{2}-
    a^{2}e^{4U}}.
    \label{omegal}
\end{equation}
This quantity is a measure for the dragging of the inertial frames 
with respect to infinity due to the rotating matter in the disk.
The dependence of $\omega_{l}$ on $\epsilon$ and $\gamma$ has been 
discussed in \cite{prd4}. As a function of $q$ it is monotonically 
decreasing as can be seen in Fig. \ref{ol}. The reason for this 
behavior is that the function $f'$ tends to zero in the limit 
$q\to 1$ for finite $\rho$, $\zeta$ whereas $g_{03}'$ changes 
shape but remains finite. Thus the overall behavior of $\omega_{l}$ 
is dominated by $f'$. The deformation of the function $g_{03}'$ 
via $q$ has however the consequence that $\omega_{l}$ has its maximum for 
large $q$ no longer at the center at the disk but near the rim.

The energy-momentum tensor at the Harrison-transformed disk can be 
calculated via (\ref{16.7}). Expressing the right-hand sides with the 
help of (\ref{f'}) and (\ref{b'}) via the original functions, we get
\begin{eqnarray}
    s_{1}' & = & 
    \frac{(1-q^{2})^{2}}{N^{2}}\left(\frac{\rho N}{2Z}b_{\rho}
    -(1-q^{4}f^{2}+q^{4}b^{2})f_{\zeta}+2q^{4}fbb_{\zeta}\right),
    \nonumber  \\
    s_{2}' & = & \frac{\rho}{2fN}\left(
    (1-q^{4}f^{2}+q^{4}b^{2})b_{\rho}+2q^{4} 
    bff_{\rho}\right),
    \nonumber  \\
    s_{3}' & = & -\frac{\rho^{3}N}{2(1-q^{2})^{2}f^{2}Z}b_{\rho}
    \label{tensor},
\end{eqnarray} 
where $N=(1-q^{2}f)^{2}+q^{4}b^{2}$.

Similarly a discontinuous electromagnetic tensor $F$ leads to a 
current density $J^{\alpha}$ 
via 
\begin{equation}
    F_{;\beta}^{\alpha\beta}=\frac{1}{\sqrt{-\det (g_{\mu\nu})}}\left(
    \sqrt{-\det (g_{\mu\nu})}F^{\alpha\beta}\right)_{,\beta}
    =-4\pi J^{\alpha}
    \label{gegen6}.
\end{equation}
Contributions to $J$ arise only 
via the normal derivatives at the disk. We define the current density 
$j^{\alpha}$ in the 
disk as $s^{\alpha\beta}$
by the relation $J^{\alpha}=:e^{-2(k-U)}j^{\alpha}\delta(\zeta)$.
The electromagnetic energy-momentum 
tensor does not produce a $\delta$-type contribution to the 
Einstein-Maxwell equations since $F$ is bounded at the disk. 
With (\ref{gegen6}) we get using  equatorial symmetry (the derivatives 
are taken at $\zeta=0^{+}$)
\begin{equation}
    2\pi j_{0}=-(\Re \Phi)_{\zeta},\quad 2\pi (j_{3}-aj_{0})
    =-\frac{\rho}{f}(
    \Im \Phi)_{\rho}
    \label{gegen7a}.
\end{equation}
This implies
\begin{eqnarray}
    2\pi j_{0} & = & \frac{q(1-q^{2})}{N^{2}}\left(-((1-q^{2}f)^{2}-
    q^{4}b^{2})f_{\zeta}+2q^{2}b(1-q^{2}f)b_{\zeta}\right)
    \nonumber  \\
    2\pi (j_{3}-aj_{0}) & = & \frac{\rho q}{(1-q^{2})Nf}\left(
    2q^{2}b(1-q^{2}f)f_{\rho}+((1-q^{2}f)^{2}-
    q^{4}b^{2})b_{\rho}\right)
    \label{gegen7b}.
\end{eqnarray}

The continuity equation (the Bianchi identity)
at the disk $T^{\mu\nu}_{;\nu}=F^{\mu\nu}
J_{\nu}=0$ leads to the condition 
\begin{equation}
	g_{00,\rho}s^{00}+2g_{03,\rho}s^{03}+g_{33,\rho}s^{33}=2(
	F_{10}j^{0}+F_{13}j^{3}).
	\label{vac20}
\end{equation}
We remark that 
one can substitute one of the equations (\ref{16.7}) by (\ref{vac20})
in the 
same way as one replaces one of the field equations by the covariant 
conservation of the energy momentum tensor in the case of 
three-dimensional perfect fluids. This makes it possible to eliminate 
$k_{\zeta}$ from (\ref{16.7}) and to treat the energy-momentum tensor at 
the disk purely on the level of the Ernst equation. It is 
straight forward to check 
the consistency of this approach with the help of (\ref{k}). Thus one 
can solve boundary value problems of dust disks without using $k$.

\begin{figure}[htb]
    \centering
    \epsfig{file=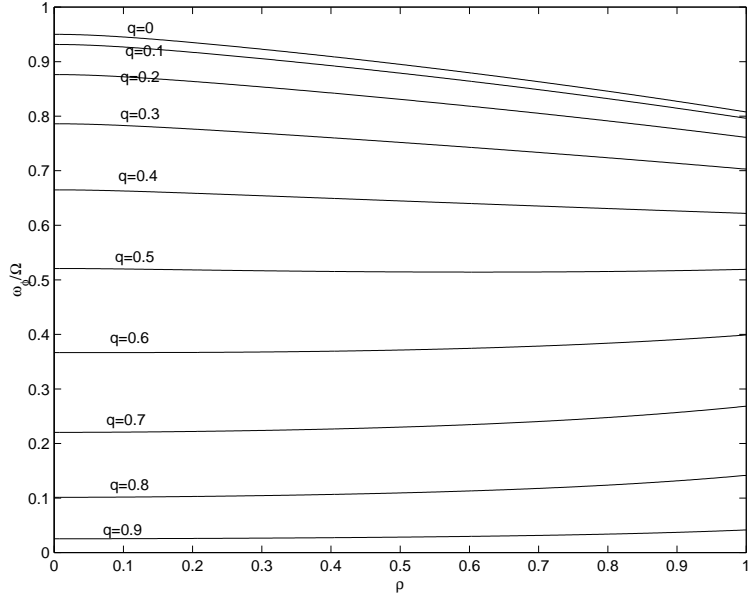,width=10cm}
    \caption{Angular velocity of the FIOs
    with respect to infinity for $\epsilon=0.85$, 
    $\gamma=0.95$, and $q=0,\ldots, 0.9$.}
    \label{of}
\end{figure}

We will interpret the matter in the disk as in \cite{zofka}
in two ways. One possible approach is to introduce 
observers for whom the tensor $s^{\alpha\beta}$ is diagonal
($\phi$-isotropic observers or FIOs) , i.e.\ we 
write $s^{\alpha\beta}$ in the form
\begin{equation}
    s^{\alpha\beta}=\sigma^{*}v^{\alpha}v^{\beta}+p^{*}w^{\alpha}w^{\beta}
    \label{fio},
\end{equation}
where $v$ and $w$ are the unit timelike respectively spacelike 
vectors $N_{v}(1,0,\omega_{\phi})$ and $N_{w}(\kappa,0,1)$ with 
$v_{\alpha}w^{\alpha}=0$. We get 
\begin{equation}
    \omega_{\phi}=\frac{g_{33}s_{00}-g_{00}s_{33}+
    \sqrt{(g_{33}s_{00}-g_{00}s_{33})^{2}+4(g_{03}s_{00}-g_{00}s_{03})
    (g_{03}s_{33}-g_{33}s_{03})}}{2(g_{03}s_{33}-g_{33}s_{03})}
    \label{omegaf}
\end{equation}
and 
\begin{equation}
    \kappa= 
    -\frac{g_{03}+\omega_{\phi}g_{33}}{g_{00}+\omega_{\phi}g_{03}}
    \label{kappa}.
\end{equation}
We show $\omega_{\phi}$ for several values of $q$ in Fig. \ref{of}. It 
can be seen that $\omega_{\phi}$ decreases monotonically with $q$. For 
large $q$ the maximum of the angular velocity is near or at the rim of 
the disk in this example.

\begin{figure}[htb]
    \centering
     \epsfig{file=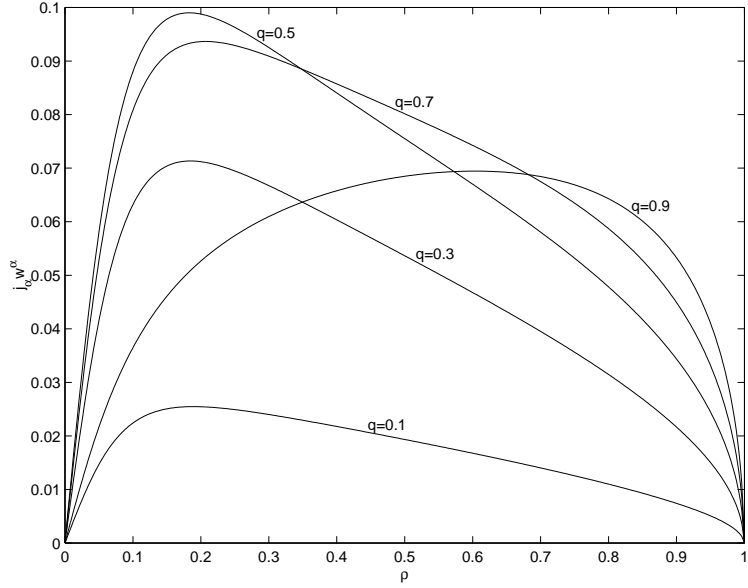,width=10cm}
    \caption{Currents in the frame of the FIOs for $\epsilon=0.85$, 
    $\gamma=0.95$, and $q=0,\ldots, 0.9$.}
    \label{j3}
\end{figure}

If the energy conditions are satisfied which is the case here, 
the FIOs can interpret 
the matter in the disk as a fluid with 
a purely azimuthal pressure. Alternatively they can interpret it as 
being made up of two streams of counter-rotating pressureless matter, 
$S^{\alpha\beta}=\frac{1}{2}\sigma^{*}(U^{\alpha}_{+}U^{\beta}_{+}+
U^{\alpha}_{-}U^{\beta}_{-})$ with 
$(U_{\pm}^{\alpha})=N(1,0,\pm\Omega_{c})$. The velocity 
$\Omega_{c}\rho$ is here below 1 (the speed of light). 
The matter streams will however not move on electro-geodesics unless 
$j_{\alpha}w^{\alpha}=0$, i.e.\ if there are no currents in the frame 
of the FIOs. However it can be seen in Fig.\ref{j3} that there are 
in general currents in the frame of the FIOs. Thus an interpretation 
of the matter by the FIOs as freely moving charged particles is 
not possible.

A different approach is to 
interpret the matter in the disk as two streams of 
particles moving on electro-geodesics in the asymptotically 
non-rotating frame. To this end we make the ansatz 
\begin{equation}
    s^{\alpha\beta}=\sigma_{m}^{+}U^{\alpha}_{+}U^{\beta}_{+}+
    \sigma_{m}^{-}U^{\alpha}_{-}U^{\beta}_{-},\quad
    j^{\alpha}=\sigma_{e}^{+}U^{\alpha}_{+}+
    \sigma_{e}^{-}U^{\alpha}_{-}
    \label{fio2}
\end{equation}
with $(U^{\alpha}_{\pm})=N_{\pm}
(1,0,\omega_{\pm})$. The angular velocity 
follows from the electro-geodesic equation for each component,
\begin{equation}
    \frac{1}{2}\sigma_{m}^{\pm}(g_{00,\rho}+2g_{03,\rho}\omega_{\pm}+
    g_{33,\rho}\omega_{\pm}^{2})=
    \sigma_{e}^{\pm}(A_{0,\rho}+A_{3,\rho}\omega_{\pm}).
    \label{fio3}
\end{equation} 
Such an interpretation is possible if the angular velocities are 
real. Moreover the velocities $\omega_{\pm}\rho$
in the disk should be smaller than 1 to avoid superluminal velocities,
and the energy densities have to be positive, 
which means we are not interested in tachyonic or otherwise exotic
matter. Relation (\ref{fio2}) leads to 
\begin{equation}
    \sigma_{e}^{+}N_{+}^{2}
    =\frac{j^{3}-\omega_{-}j^{0}}{\omega_{+}-\omega_{-}},
    \quad \sigma_{e}^{-}N_{-}^{2}=
    \frac{j^{3}-\omega_{+}j^{0}}{\omega_{-}-\omega_{+}}
    \label{fio4},
\end{equation}
\begin{equation}
    \sigma_{m}^{+}N_{+}^{2}=\frac{s^{03}-\omega_{-}s^{00}}{
    \omega_{+}-\omega_{-}},\quad
    \sigma_{m}^{-}N_{-}^{2}=\frac{s^{03}-\omega_{+}s^{00}}{
    \omega_{-}-\omega_{+}}
    \label{fio5},
\end{equation}
and
\begin{equation}
    \omega_{-}=\frac{s^{33}-\omega_{+}s^{03}}{s^{03}-\omega_{+}s^{00}}
    \label{fio6}.
\end{equation}
If we enter (\ref{fio3}) with this, we obtain
\begin{equation}
    \omega_{\pm}=\frac{-T_{2}\pm\sqrt{T_{2}^{2}-T_{1}T_{3}}}{T_{1}}
    \label{fio7}
\end{equation}
with 
\begin{eqnarray}
    T_{1} & = & g_{33,\rho}-2A_{3,\rho}\frac{j^{0}s^{03}-j^{3}s^{00}}{
    s^{03}s^{03}-s^{00}s^{33}}
    \nonumber  \\
    T_{2} & = & g_{03,\rho}-A_{0,\rho}\frac{j^{0}s^{03}-j^{3}s^{00}}{
    s^{03}s^{03}-s^{00}s^{33}}-A_{3,\rho}\frac{j^{3}s^{03}-j^{0}s^{33}}{
    s^{03}s^{03}-s^{00}s^{33}}
    \nonumber  \\
    T_{3} & = & g_{00,\rho}-2A_{0,\rho}\frac{j^{3}s^{03}-j^{0}s^{33}}{
    s^{03}s^{03}-s^{00}s^{33}}
    \label{fio8}.
\end{eqnarray}
The densities then follow from (\ref{fio2}) where the continuity 
equation (\ref{vac20}) guarantees that this system can be 
solved. 

\begin{figure}[htb]
    \centering
    \epsfig{file=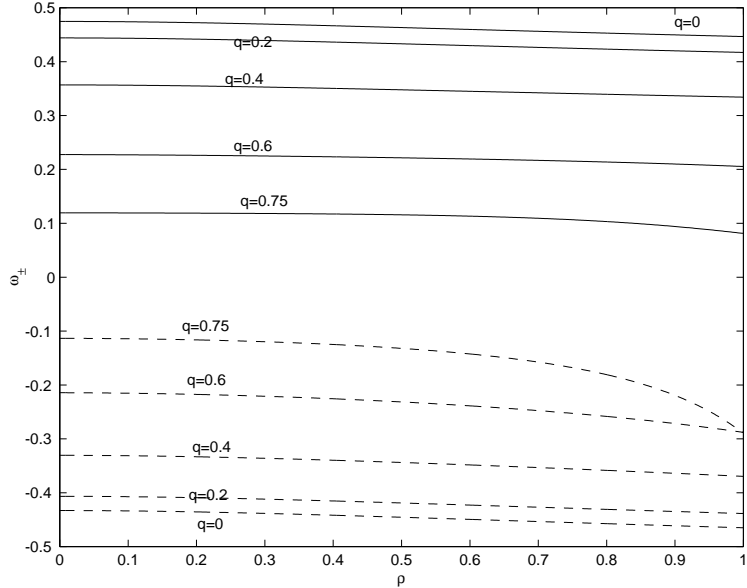,width=10cm}
    \caption{Angular velocities $\omega_{\pm}$ for $\epsilon=0.36$, 
    $\gamma=0.08$, and $q=0,0.2,\ldots,0.8$.}
    \label{omp}
\end{figure}

\begin{figure}[htb]
    \centering
     \epsfig{file=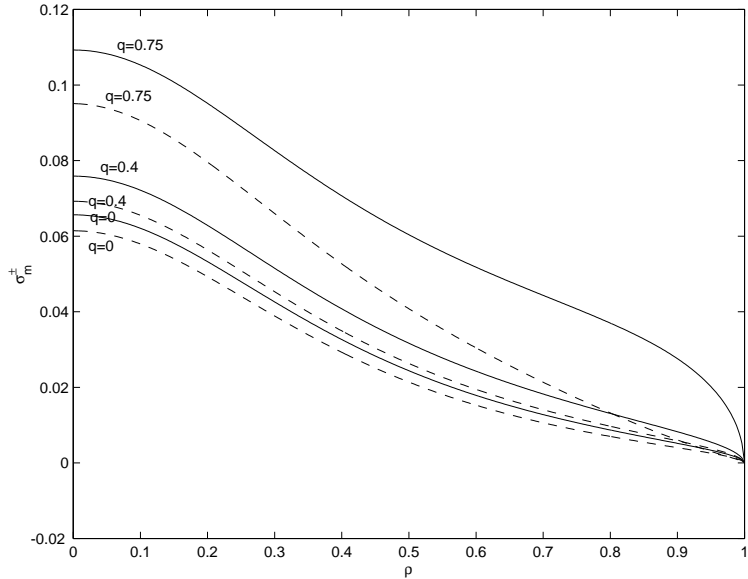,width=10cm}
    \caption{Matter densities  $\sigma_{m}^{+}$ and $\sigma_{m}^{-}$ 
    (dashed) for $\epsilon=0.36$, 
    $\gamma=0.08$, and $q=0,0.4,0.8$.}
    \label{smpm}
\end{figure}

\begin{figure}[htb]
    \centering
    \epsfig{file=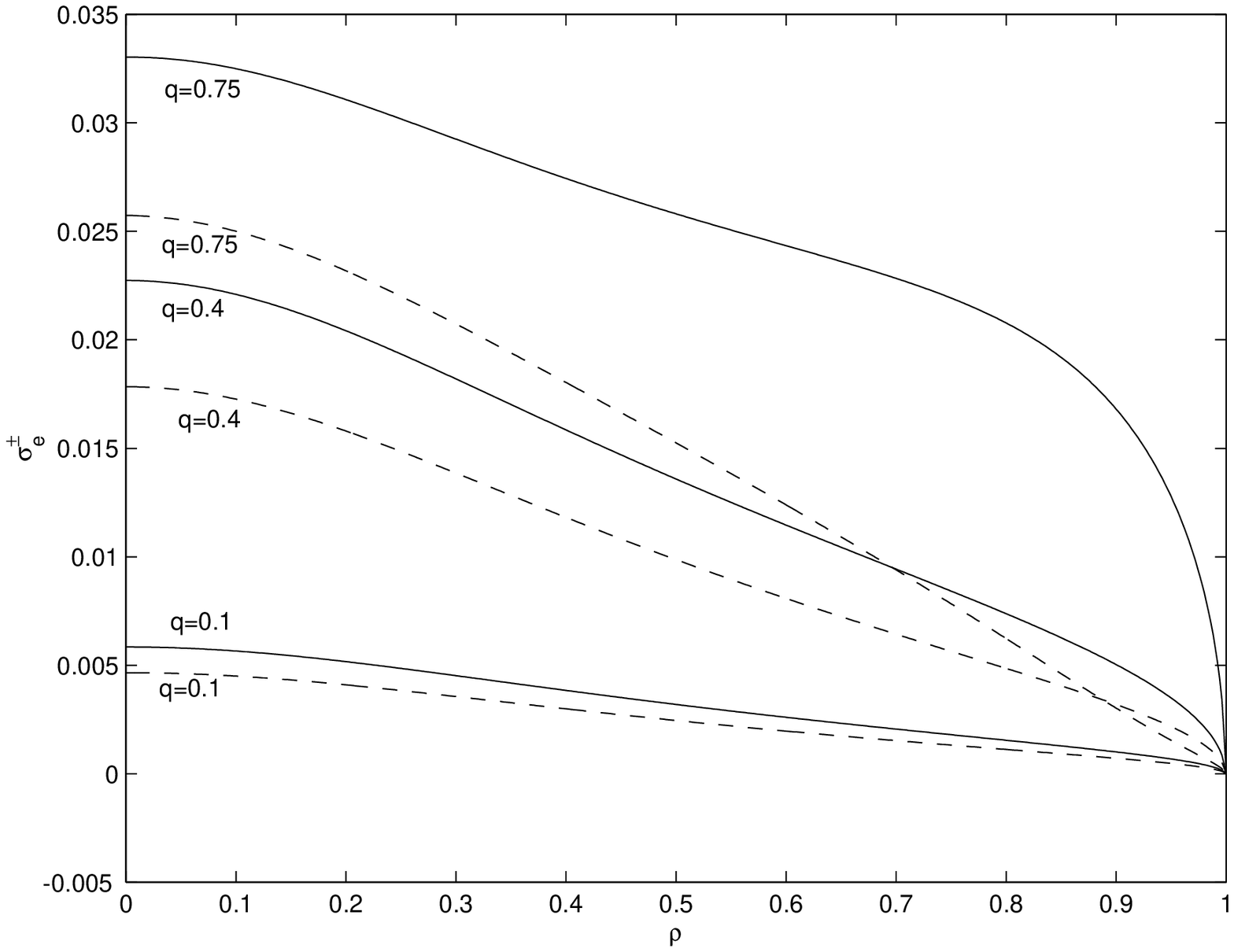,width=10cm}
    \caption{Charge densities  $\sigma_{e}^{+}$ and $\sigma_{e}^{-}$ 
    (dashed) for $\epsilon=0.36$, 
    $\gamma=0.08$, and $q=0,0.4,0.8$.}
    \label{sepm}
\end{figure}

Numerically one finds that the angular velocities are real, but in a 
wide range of the parameters there are negative energy densities and 
tachyonic behavior. Already in 
the uncharged case there are infinite velocities in strongly 
relativistic settings with negligible counter-rotation which are due 
to extrema of the metric function $g_{33}$ in the disk. In this case 
the quantity $T_{1}$ in (\ref{fio8}) is zero which leads to a 
diverging $\omega_{-}$. Increasing $q$ only enhances this effect. The 
result is that an interpretation as non-tachyonic counter-rotating 
matter on electro-geodesics with positive energy densities 
is only possible if $q$, $\epsilon$ and 
$\gamma$ are not too large. In other words large values of $q$ are in 
this setting only possible in post-Newtonian or nearly static 
situations. We show plots of the angular velocities $\omega_{\pm}$ 
in Fig.~\ref{omp}
for $\epsilon=0.36$ and $\gamma=.08$ where values of $q$ up to $0.75$ 
are possible.
The corresponding densities are given in Fig. \ref{smpm},
and the charge densities in Fig. \ref{sepm}.
The densities vanish always at the rim of the disk, the 
charge densities are identically zero for $q=0$. For larger values of 
$q$, the angular velocity $\omega_{-}$ becomes bigger and bigger
at the rim of 
the disk and finally diverges. The density $\sigma_{m}^{-}$ is in 
this case negative in the vicinity of the rim of the disk.

\section{Conclusion}
In this article we discussed Harrison transformations of 
solutions to the stationary axisymmetric Einstein equations in vacuum 
to generate solutions to the corresponding Einstein-Maxwell 
equations. The case of hyperelliptic solutions to the Ernst equation 
was studied in detail. This led to a class of solutions to the 
Einstein-Maxwell system which contains the following degrees of 
freedom: a real-valued function $G$, a set of complex branch points 
$E_{i}$ and a parameter $q$ related to the total charge of the 
spacetime. Since the complete metric is given explicitly, 
one could in principle try to solve physically interesting 
boundary value problems as dust disks with a magnetic field within 
this class.

As a first step in this direction we have considered a Harrison 
transformation of a family of counter-rotating dust 
disk solutions where the matter rotates in the frame of the FIOs on 
geodesics. The matter in the disk could be interpreted as moving on 
electro-geodesics in a certain range of the physical parameters, but 
in general not in the frame of the FIOs. 
As expected one cannot hope that a Harrison transformation preserves 
certain features of the matter 
as electro-geodesic motion in the FIO frame. 
If one wants to generate solutions with such properties or 
disks made up of charged pressureless particles without additional 
currents, one will have to solve the corresponding 
boundary value problem. Whether this will be possible within the 
class of Harrison transformed hyperelliptic solutions of the Ernst 
equation is an open question. For more general cases one will have to 
use Korotkin's theta functional solutions to the Einstein-Maxwell 
equations on three-sheeted surfaces, where however
the powerful calculus of hyperelliptic surfaces can no longer be 
used. 

\textbf{Acknowledgement:} I thank D.~Maison, who interested me in the 
subject, for many helpful discussions and hints. This work was 
supported by the Schloessmann foundation.

\end{document}